\documentclass{appolb}
\usepackage{graphicx}
\usepackage{eqnarray}

\newcommand{\EPA}{\mathcal{E}}   
\newcommand{\KEA}{\mathcal{T}}   
\newcommand{\VEA}{\mathcal{V}}   

\begin{document}
\title{Nuclear Energy Density Functional for KIDS%
\thanks{Presented at the Zakopane Conference on Nuclear Physics “Extremes of the Nuclear
Landscape”, Zakopane, Poland, August 28 $-$ September 4, 2016}%
}
\author{
       Hana Gil$^1$,  Panagiota Papakonstantinou$^2$,  Chang Ho Hyun$^3$, Tae-Sun Park$^4$,  
Yongseok Oh$^1$ 
\address{$^1$Department of Physics, Kyungpook National University, Daegu 41566, Korea\\
           $^2$Rare Isotope Science Project, Institute for Basic Science, Daejeon 34047, Korea\\
          $^3$Department of Physics Education, Daegu University, Gyeongsan 38453, Korea \\ 
       $^4$Department of Physics, Sungkyunkwan University, Suwon 16419, Korea \\ 
              }
           }
\maketitle
\begin{abstract}
The density functional theory (DFT) 
is based on the existence and uniqueness of a universal functional $E[\rho]$, which determines the dependence of the total energy on single-particle density distributions. However, DFT says nothing about the form of the functional. 
Our strategy is to first look at what we know, from independent considerations, about the analytical density dependence of the energy of nuclear matter and then, for practical applications, to obtain an appropriate  density-dependent effective interaction by reverse engineering. In a previous work on homogeneous matter, we identified the most essential terms to include in our ``KIDS" functional, named after the early-stage participating institutes. We now  present  first results for finite nuclei, namely the energies and radii of $^{16,28}$O, $^{40,60}$Ca.  
\end{abstract}


%
%
%
%
\section{Introduction} 

The density functional theory (DFT) for interacting quantum many-body systems rests on the famous Hohenberg-Kohn and Kohn-Sham theorems. The fundamentals are simple: For a collection of a given number and species of particles, there is a unique universal functional $E[\rho]$, which determines how the total energy depends on single-particle density distributions. The density distributions in turn can be represented by those of an auxiliary system of non-interacting particles in some single-particle potential $V$. 

DFT, however, tells us nothing about the form of the functional and the potential, and provides no guidance to that end. In nuclear physics, one traditionally begins by assuming a specific form for the effective interaction and then tries to obtain the objects $E[\rho]$ and $V$ within the Hartree-Fock approximation. Methods beyond mean field are also explored~\cite{BHR2003}. In this project we rather do the opposite. We start from what we know about the energy of the strongly interacting Fermion system that is nuclear matter. As we argue below and in Ref.~\cite{PPL2016X}, the energy per particle in dilute nuclear matter must include low-order powers of the Fermi momentum $k_F$, or $\rho^{1/3}$, beginning with the kinetic energy term, $k_F^2$. We then obtain a density-dependent effective interaction which reproduces the desired functional within (at this stage) the self-consistent mean field approach (Hartree-Fock). 

In Sec.~\ref{S:HNM} we summarize our reasoning and findings 
for nuclear matter. In Sec.~\ref{S:Nucl} we obtain an auxiliary effective ineteraction for mean-field calculations and apply it to spin-saturated nuclei. Prospects are listed in Sec.~\ref{S:Prosp}. 

\section{\label{S:HNM}Form of $E[\rho ]$ in homogeneous matter} 

The basic idea of the present approach is to consider the Fermi momentum as the fundamental variable of the functional and write the functional as a power expansion in $k_F$ beginning with the kinetic energy contribution: 
\begin{equation}
\EPA \equiv E/A = \KEA + \sum_{i \geq 0} c_i \rho^{1+i/3} = \KEA + \VEA \, . 
\label{Eq:pow}
\end{equation}\\[-3mm]
There are two lines of physical reasoning that lead to the above form. 

In Ref.~\cite{PPL2016X} we argued that, within a wide range of densities relevant for nuclei and neutron stars, nuclear matter is dilute with respect to the range of heavy-meson exchange. 
We postulated that the average effect of the pions is a modification of the coupling constants between nucleons and heavy mesons. 
Then, we may write the energy per particle in the same form as given by an effective field theory for dilute Fermion systems~\cite{HaF2000}, namely an expansion in powers of $k_F$. The expansion coefficients far from the (trully) dilute limit presumably are related to in-medium scattering lengths, which, however, are not known. With the help of pseudodata for symmetric nuclear matter (SNM) and pure neutron matter (PNM) we showed that the lower-order terms $c_{0,1}$ are the most relevant and robust (their values change weakly as we change the details of the fits; they acquire similar values if we determine them from the saturation point; they give better fits). Logarithmic terms owing to three-nucleon forces were found irrelevant. Our first set of parameters predicts correctly results from chiral EFT and produces a neutron star mass-radius relation consistent with observations~\cite{PPL2016X}. 
The second argument comes from the Brueckner theory for strongly interacting Fermi systems. When a ``realistic" potential is assumed between nucleons, namely one characterized by a short-range repulsive core and a medium-range attractive part, $\VEA$ is precisely the sum of $k_F$ powers written in Eq.~(\ref{Eq:pow}), as an inspection of the expressions given in \cite{FeW1971} shows. 

As detailed in \cite{PPL2016X}, we determine $c_i(\delta )$ for SNM (asymmetry $\delta =0$) and PNM ($\delta =1$), using $i\leq 3$, from fits to pseudodata~\cite{APR1998}.  At present we assume for simplicity that the sloppiest parameter in SNM, namely $c_3$, vanishes. We do obtain $c_3(0)=0$ for a given cost function ($\beta =0.97$, cf. \cite{PPL2016X}). Then we only have three parameters for SNM, which we determine such that the saturation density of SNM is $\varrho_0=0.16$~fm$^{-3}$, the energy at saturation is $\EPA (\varrho_0)=-16$~MeV, and the incompressibility is $K_{\infty}=240$~MeV. We get $c_0(0)=-664.516$~fm$^3$MeV, $c_1(0)=763.545$~fm$^4$MeV, and $c_2(0)=40.133$~fm$^5$MeV, close to the fitted values in Ref.~\cite{PPL2016X}. For PNM we use the fit obtained with $\beta =0.97$, which provides $c_3(0)$=0. 

\section{\label{S:Nucl}Exploratory results in finite nuclei} 

We observe that our functional can be obtained within the Hartree-Fock approximation (self-consistent mean field) from a Skyrme-like effective interaction with the usual momentum-dependent terms with couplings $t_1,t_2$~\cite{SHFcode} and with three density-dependent terms, ($t_3+t_{x3}\hat{P}_r)\rho^{1/3}\delta (\vec{r}_{12})$, ($t_3'+t_{x3}'\hat{P}_r)\rho^{2/3}\delta (\vec{r}_{12})$, ($t_3{''}+t_{x3}{''}\hat{P}_r)\rho\delta (\vec{r}_{12})$, where the newly defined $t_{xi}\equiv t_ix_i$ can be finite even if $t_i$ vanish. Thus we can transform it into a Skyrme-type functional and apply it to finite nuclei. As long as the spin degree of freedom is not probed, the procedure is legitimate within DFT.  
\begin{figure}[h]
\includegraphics[height=5.0cm,width=12cm]{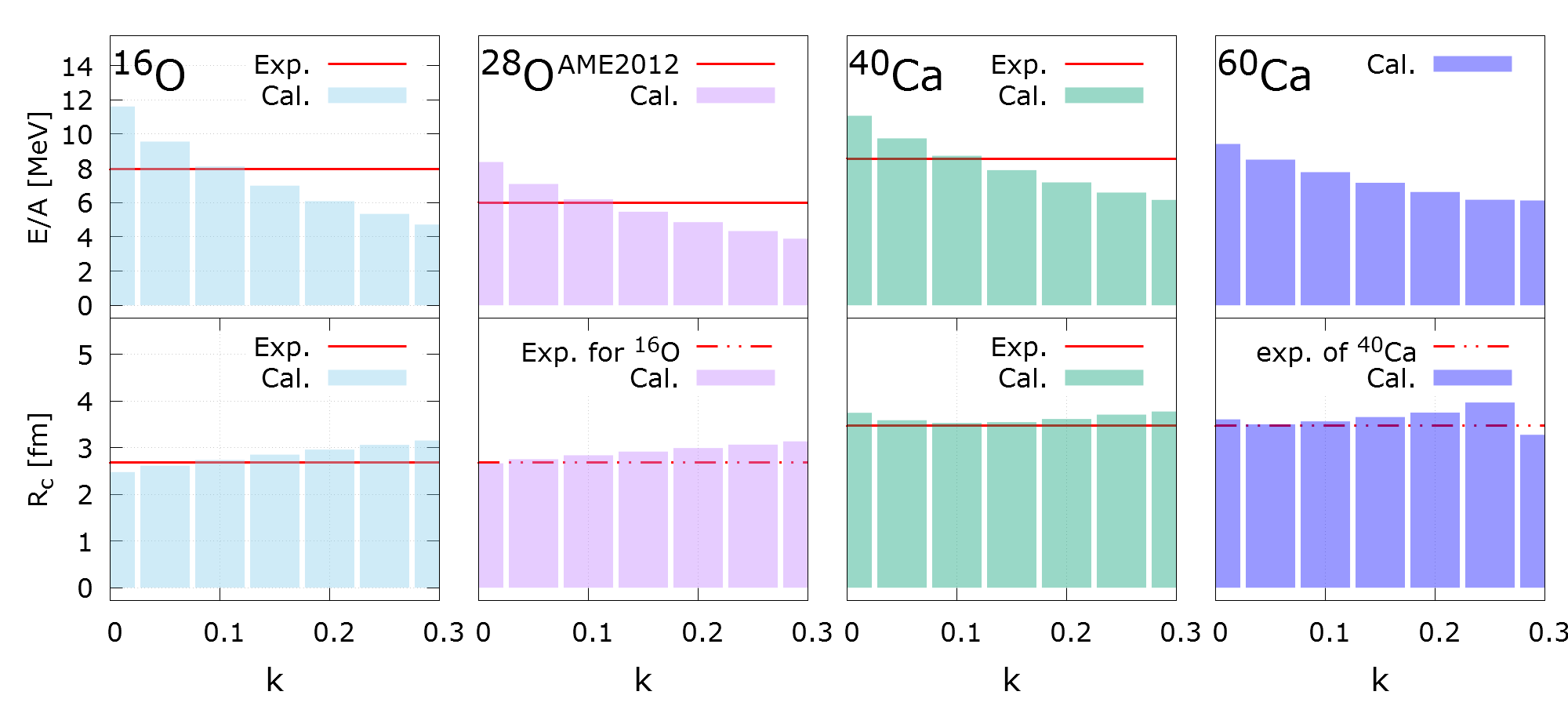} 
\caption{\label{fig}Energy per particle [MeV] and charge radii of spin-saturated nuclei as a function of the parameter $k$. 
Full lines: AME2012 values. Dot-dashed lines: Charge radius of $^{16}$O (in $^{28}$O graph) or $^{40}$Ca (in $^{60}$Ca graph).}
\end{figure} 
The contribution of the momentum-dependent terms and of the $t_3'$ term in homogeneous matter should add up to the $c_2\rho^{5/3}$ terms. We thus write $c_2=kc_2 + (1-k)c_2 \equiv c_2^{t_1,t_2} + c_2^{t_3'} $. The parameter $k$ is unknown and will be determined from the properties of finite nuclei. At this exploratory stage we consider $k$ independent of the asymmetry. 
The $t_i$ and $x_i$ parameters are straightforward to obtain from the functional form and $c_i$, see, e.g., \cite{Dob2016}, though we cannot constrain $x_1$ and $x_2$ without the spin degree of freedom.  At this stage we set $x_1=x_2=0$. 
The above procedure gives: 
\[ 
t_0 = -1772.04\,	\mathrm{fm}^3\mathrm{MeV} ; \, 
t_1 =  2492.11 \times k	  \,  \mathrm{fm}^5\mathrm{MeV} ; \,
t_2 = -1459.77 \times k   \,  	\mathrm{fm}^5\mathrm{MeV}  
\] \vspace{-7mm} 
\[ 
t_3 = 12216.71			\,\mathrm{fm}^4\mathrm{MeV} ; \quad 
t_3' = 642.13\times (1-k) \,	\mathrm{fm}^5\mathrm{MeV} ; \quad 
t_3'' = 0.00				\,\mathrm{fm}^6\mathrm{MeV}
\]
and $t_{x0}=-127.59$~fm$^3$MeV, $t_{x3}=-11972.38$~fm$^4$MeV, $t_{x3}'=33153.17$~fm$^5$MeV, $t_{x3}''=-22955.28$~fm$^6$MeV. 
Using a Skyrme-Hartree-Fock code~\cite{SHFcode}
we calculate the energy and radius of the spin-saturated nuclei $^{16}$O, $^{28}$O, $^{40}$Ca, $^{60}$Ca. The results are summarized in Fig. 1. For a value of $k\approx 0.1$, corresponding to an almost bare-nucleon effective mass, all properties are well reproduced. 
The results provide a first ``proof of principle" for our approach. 

\section{\label{S:Prosp}Prospects} 

There are several paths to pursue next: 
The analysis of nuclear matter may be repeated with different higher-order terms, ($\rho^{7/2}$, or $\rho^{8/3}$) instead of $\rho^2$.  
We need to determine the spin dependence 
ideally via pseudodata for polarized matter, and a spin-orbit term. 
Data for finite nuclei can be used to fine-tune the parameters. 
We also wish to study excitations within RPA. 

{~}\\
\noindent 
{\bf Acknowledgements} Work supported by: the Rare Isotope Science Project of the Institute for Basic Science funded by Ministry of Science, ICT and Future Planning and the 
National Research Foundation (NRF) of Korea (2013M7A1A1075764)
and 
and the Basic Science Research Program through the NRF funded by the Ministry of 
Education (NRF- 2013R1A1A2063824, NRF-2014R1A1A2054096, NRF-2015R1D1A1A01059603).


\begin{thebibliography}{1}

\bibitem{BHR2003}
M.~Bender, P.-H. Heenen, P.-G. Reinhard,
\newblock {\em Rev. Mod. Phys.}
\newblock {\bf  75}, 121 (2003).

\bibitem{PPL2016X}
P.~Papakonstantinou, {T.-S.} Park, Y.~Lim, {C.H.} Hyun,
\newblock {\em arXiv:}
\newblock {\bf  1606.04219} (2016).

\bibitem{HaF2000}
H.-W. Hammer, R.J. Furnstahl,
\newblock {\em Nucl. Phys. A}
\newblock {\bf  678}, 277  (2000).

\bibitem{FeW1971}
{A.L.} Fetter, {J.D.} Walecka,
\newblock {\em {Quantum Theory of Many-Particle Systems}},
\newblock Dover Publications,
\newblock New York 1971.

\bibitem{APR1998}
A.~Akmal, V.~R. Pandharipande, D.~G. Ravenhall,
\newblock {\em Phys. Rev. C}
\newblock {\bf  58}, 1804 (1998).

\bibitem{SHFcode}
P.-G. Reinhard,
\newblock
\newblock {\em {Computational Nuclear Physics I - Nuclear Structure}} (ed. K.
  Langanke, J.E. Maruhn and S.E. Koonin, Springer, New York 1991), {\em p.28}.

\bibitem{Dob2016}
J.~Dobaczewski,
\newblock {\em J. Phys. G: Nucl. Part. Phys.}
\newblock {\bf  43}, 04LT01 (2016).

\end{thebibliography}
\end{document}